\documentclass{elsart}
\usepackage{natbib}
\usepackage{graphics}
\usepackage{longtable}

\begin{document}
\newcommand{\xbeq}{\begin{eqnarray}} \newcommand{\xeeq}{\end{eqnarray}}

\runauthor{}
\begin{frontmatter}
  \title{MC3D - 3D Continuum Radiative Transfer\\Version 2}
  \author[Caltech]{S.\ Wolf}
 
  \address[Caltech]{California Institute of Technology, 1200 E California Blvd, 
	Mail Code 220-6, Pasadena CA 91125,
    	USA, swolf@ipac.caltech.edu}
  
  \begin{abstract}
A revised and greatly improved version of the 3D continuum radiative transfer code
MC3D is presented.
It is based on the Monte-Carlo method and solves the radiative transfer problem 
self-consistently. It is designed for the simulation of dust temperatures in arbitrary
geometric configurations and the resulting observables: spectral energy distributions,
wavelength-dependent images, and polarization maps. The main objective is the investigation
of ``dust-dominated'' astrophysical systems such as young stellar objects 
surrounded by an optically thick circumstellar
disk and an optically thin(ner) envelope, debris disks around more evolved stars, 
asymptotic giant branch stars,
the dust component of the interstellar medium, and active galactic nuclei.

\vspace{+5mm}
\hspace*{-4mm}{\sl PACS:}
95.30.Jx -- Radiative transfer\\
  \end{abstract}

\end{frontmatter}

\section{Introduction}

The 3D continuum radiative transfer (RT) code, MC3D, combines
the most recent Monte Carlo (MC) radiative transfer concepts for both
the self-consistent RT, i.e., the estimation of spatial dust temperature distributions,
and pure scattering applications, taking into account the polarization state
of the radiation field. It has been tested intensively and compared with
grid-based and MC RT codes 
(see, e.g., Pascucci et al.~\cite{pa02}, Wolf~\cite{myphd}).

\begin{figure}[t]
 \resizebox{\hsize}{!}{\includegraphics{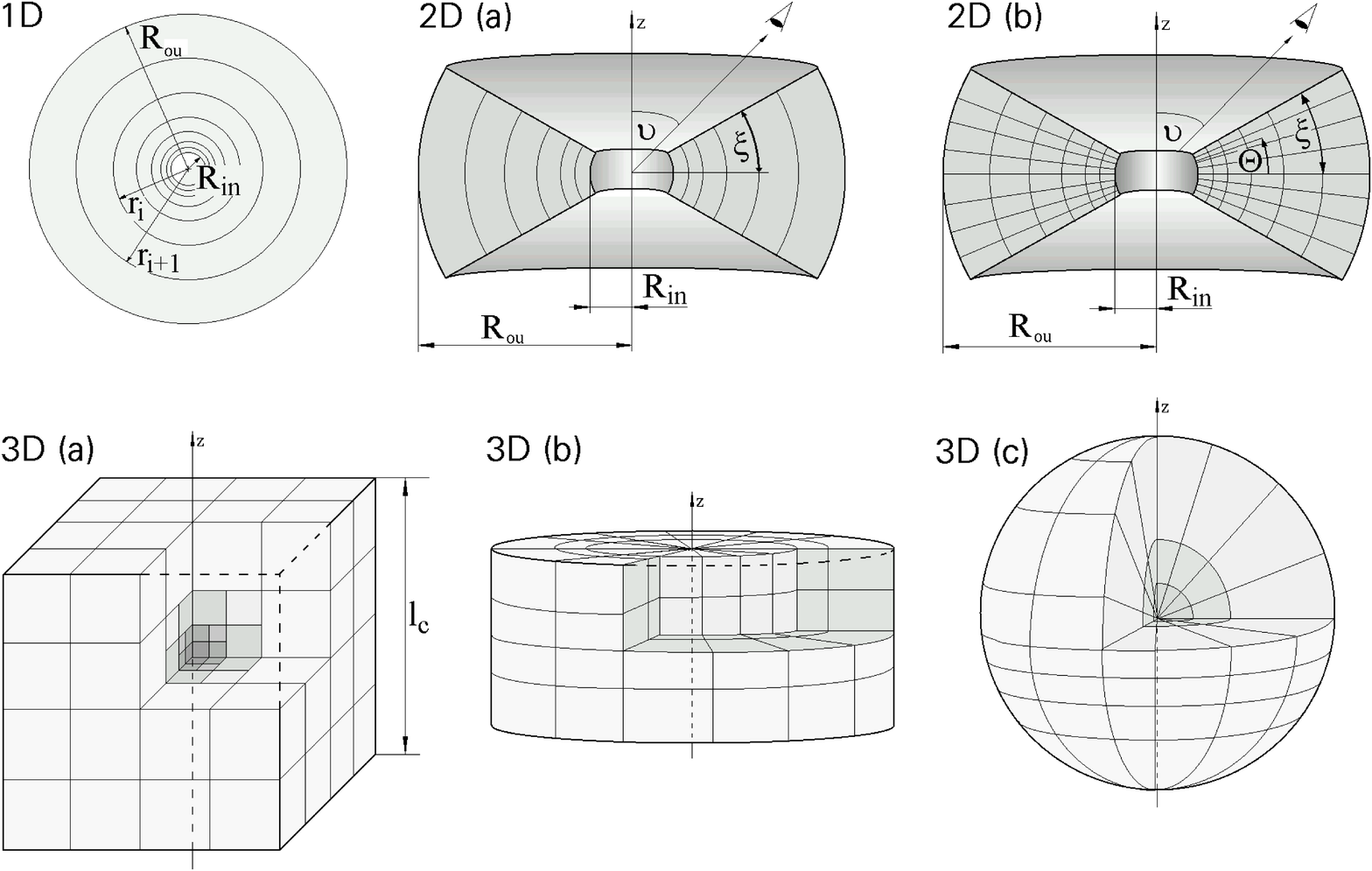}} 
 \caption{Model geometries: 
Examples for the subdivision of the model space into energy
storage cells (see Sect.~\ref{esc} for further explanations). 
{\bf 1D:} radial symmetry;
{\bf 2D(a):} radial symmetry inside a disk;
{\bf 2D(b):} fully two-dimensional model for configurations with a radial 
and vertical dependence of the density distribution;
{\bf 3D(a-c):} three-dimensional models considered in cartesian, cylindrical, 
and spherical coordinates. }
 \label{gridex}
\end{figure}

In addition to the previous version of MC3D (Wolf \& Henning~\cite{wolf00}; 
see also Wolf et al.~\cite{wolf99}),
MC3D\,(V2) contains several improvements which allow a much more efficient solution
of the radiative transfer problem in three-dimensional, arbitrary dust configurations.
The main features are:
\begin{itemize}
\item A new concept for {\em estimation of dust temperature distributions}
which was first described by Bjorkman \& Wood~\cite{bj01} in the context of the solution
of the radiative transfer problem in circumstellar disks. It is based on the immediate
correction of the dust grain temperature and the reemission radiation field 
after absorption of a photon packet.
Beyond a faster estimation of the dust equilibrium temperature without iterations,
it allows the application of the code for optical depths far above the previous
$\tau \approx 10^3$ limit.
\item A new concept for the {\em test photon transfer}, based on the work of Lucy~\cite{lu99},
(previously developed for one-dimensional model geometries), which takes into account the absorption 
not only at the points of interaction of the test photons with the dust grains but also 
in between. Therefore, the temperature correction can be performed in all volume elements
on the test photon path and consequently, there is no more lower optical depth limit
for the application of MC3D.
\item A new concept for the simulation of scattering in optically thin dust/electron
configurations which has been firstly described by Cashwell \& Everett~\cite{ca59}
(see also Witt~\cite{wi77} and M$\acute{\rm e}$nard~\cite{men88} for former applications in
one- and two-dimensional model geometries) which makes use 
of the {\em enforced scattering method}.
\item The simulation of the RT in {\em dust grain mixtures}
consisting of grains with different chemical composition and size.
\item A {\em raytracer} for the simulation of images and spectral energy distributions (SEDs)
which may be used for the simulation of the dust reemission in the 
midinfrared-millimeter wavelength range, where scattering is of minor importance.
\item An accelerated solution of the temperature estimation in flat disks
with or without vertical density structure.
\end{itemize}
These features have been adapted to and tested for several 1D, 2D, and 3D model
geometries which are shown in Fig.~\ref{gridex}.

Previous applications of MC3D cover
feasibility studies of extrasolar planet detections (Wolf et al.~\cite{wolf02b}),
the RT in the clumpy circumstellar environment of young stellar objects (Wolf et al.~\cite{wolf98}),
polarization studies of T~Tauri stars (Wolf et al.~\cite{wolf01}),
AGN polarization models (Wolf \& Henning~\cite{wolfh99}), 
a solution for the multiple scattering of polarized radiation by non-spherical grains 
(Wolf et al.~\cite{wovo02}),
and the inverse RT based on the MC method (Wolf~\cite{wolfir01}).

The executables of MC3D\,(V2) can be downloaded for several model geometries
and computer platforms
from \\
{\tt http://www.mpia-hd.mpg.de/FRINGE/SOFTWARE/mc3d/}\\ 
(current US mirror page:\\ 
{\tt http://spider.ipac.caltech.edu/staff/swolf/mc3d/}).\\
Furthermore,
a detailed description, usage instructions, and IDL\footnote{Interactive Data Language}
routines for a subsequent data analysis of the code are provided there.
Very frequently  used model geometries, such as disks with a radial and
vertical variable density distribution, which is often used for the analysis
of SEDs, images and polarization maps of circumstellar disks (see, e.g.,
Wood et al.~\cite{wood02}, Cotera et al.~\cite{cot01}, Fischer et al.~\cite{fi96}), 
are available.
However, the author encourages the reader to ask for further model geometries.

In Sect.~\ref{general} the general scheme for the solution of the RT problem
on the basis of the MC method as it is applied in MC3D 
is briefly outlined. In Sect.~\ref{concepts} the two different concepts
for the estimation of the spatial dust temperature distribution 
(iterative radiative transfer and immediate reemission) are discussed.
In the following Sect.~\ref{obs}, the determination of the observables is described,
and Sect.~\ref{acc} introduces further concepts which are
used in MC3D in order to increase the efficiency of the RT algorithms.

\section{General remarks about the applied solution 
of the radiative transfer problem}\label{general}

Each model to be considered with MC3D consists of two independently defined components:
\begin{enumerate}
\item The spatial density distribution of the scattering and absorbing medium, and
\item The radiative source(s).
\end{enumerate}
Both must be defined inside a convex model space in order to exclude
the case of radiation leaving the model space at one point and entering
it at another.

\subsection{Scattering and absorbing medium}

MC3D handles spherically symmetric scatterers/absorbers, such as spherical dust grains
and electrons (in the Thomson scattering regime). The required optical properties,
i.e., the absorption and extinction efficiency ($Q_{\rm abs}$, $Q_{\rm ext}$), 
the special Mueller matrix $\hat{S}$ (see Sect.~\ref{ptrsca}, App.~\ref{scattmat}), 
and quantities to be derived from these -- 
the albedo $A$,
the extinction and scattering cross section ($C_{\rm ext}$, $Q_{\rm sca}$), and 
the scattering function --
can be calculated with an embedded Mie scattering routine (using the Mie scattering
algorithm published by Bohren \& Huffman~\cite{bo83}) on the basis of the real and imaginary
refractive index ($n$,$k$) and the particle radius $a$.
Since Thomson scattering can be -- formally -- considered as a special form
of the Mie scattering formalism (no absorption, no wavelength dependence), in
the following only dust grains are considered as the scattering and absorbing
medium, covering also the special case of electrons.

MC3D allows simulation of the RT in arbitrary density distributions
$\rho_j(\vec{r})$, where $\vec{r}$ represents the spatial coordinate and 
the index $j$ refers to the considered dust species (characterized by the grain size
and chemical composition). If MC3D is used to simulate the {\em scattering} by dust grains
only (for instance for the simulation of images, polarization maps or SEDs in 
the visual-midinfrared wavelength range), the mean optical properties
of dust grain mixtures consisting of 3 chemically different components
and a size distribution following a power-law $n(a) \propto a^{-\alpha}$
($n(a)$ is the number of dust grains with the radius $a$; $\alpha$=const.)
can be estimated on the basis of the formalism being described
in Appendix~\ref{mix}. As found by Wolf~\cite{wolf02c}, this concept of mean
dust grains also represents a reasonable approximation for the estimation of temperature
distributions in optically thick density distributions and the simulation of SEDs
in general.

The density distribution may be defined either analytically or on a predefined
grid  in order to allow the implementation of density distributions resulting
from hydrodynamical simulations (for examples, see Wolf et al.~\cite{wolf99}, 
Sect.~4; Wolf et al.~\cite{wolf02b}).

\subsection{Radiation sources}\label{rasou}

\begin{figure}[t]
 \resizebox{\hsize}{!}{\includegraphics{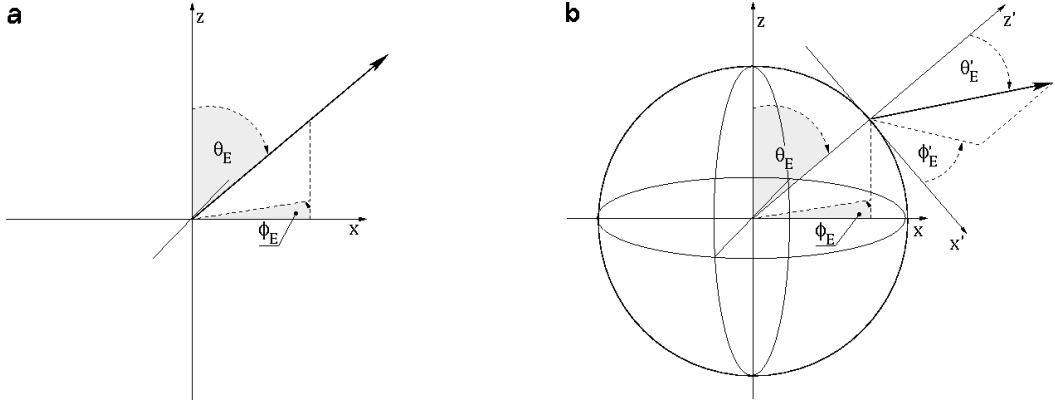}} 
 \caption{Emission angles. [a] Pointlike star, [b] Extended source.}
 \label{emiss}
\end{figure}

In principle, radiation sources, that are arbitrary with respect to their number and
spatial configuration, spatial extent, intrinsic SED, and radiation characteristic
can be considered (see Wolf et al.~\cite{wolf99}, \cite{wolf02b} for examples).
In MC3D the RT is simulated at certain wavelengths within the wavelength
range [$\lambda_{\rm min}$, $\lambda_{\rm max}$]. For this reason, the monochromatic
luminosity\footnote{Between the (bolometric) luminosity $L$ and the monochromatic luminosity
$L_{\lambda}$ the following relation exists: $L=\int_0^{\infty} L_\lambda\, {\rm d}\lambda$.} 
$L_{\lambda}$ of the radiation source(s) is partitioned into $n_{\rm Photon}$ so-called
``weighted photons'' (Sobolev~\cite{so82}) each of which is characterized
by its wavelength $\lambda$ and Stokes vector $\hat{I}=(I,Q,U,V)^{\rm T}$.
The polarization state of a photon is determined
by the Stokes vector components as follows:
\begin{equation}\label{polsdsgbt}
  \begin{array}{lcl}
  {\mbox {\rm (a) Linear polarization degree               }} & 
  {\rm :} & P_{\rm l} = \frac{\sqrt{(Q^2+U^2)}}{I}\\ 
  {\mbox {\rm (b) Orientation of the linear polarization}} & 
  {\rm :} & \gamma = \frac{1}{2} \arctan\left({\frac{U}{Q}}\right)\\
  {\mbox {\rm (c) Circular polarization degree             }} & 
  {\rm :} & P_{\rm c} = \frac{V}{I}\\
  \end{array}
\end{equation}
The two main radiation sources in MC3D RT simulations shall be introduced in brief:
\begin{enumerate}
\item Pointlike Star: Here the emission direction of photons, 
described in spherical coordinates ($\theta_{\rm E}$, $\phi_{\rm E}$) 
is given by
	\begin{equation}\label{pointlike}
	\cos\theta_{\rm E} = -1 + 2\, Z_1, \hspace*{1cm} 
	\phi_{\rm E} = 2\,\pi\,Z_2,\
	\end{equation}
\item Extended Star
(radius $R_{S}$, 
radiation characteristic at each point on the stellar surface: $I \propto \cos(\theta_{\rm E}')$):
	\begin{equation}\label{extended}
        \sin\theta_{\rm E}' = \sqrt{Z_3}, \hspace*{1cm} 
        \phi_{\rm E}' = 2\,\pi\,Z_4\ ,
	\end{equation}
\end{enumerate}
whereby $\theta_{\rm E}'$ and $\phi_{\rm E}'$ are related to the $z'$ axis
parallel to the radius vector ($R_{s},\theta_{\rm E},\phi_{\rm E}$) which
can be determined using Eq.~\ref{pointlike} (see Fig.~\ref{emiss} for explanation).
Here and in the following, $Z_i$ ($i=1,2,\ldots$) are random numbers uniformly 
distributed in the interval
$[0,1]$. For their determination a random number generator, which minimizes the
sequential correlations between random numbers by the combination of different random
number generators, is used (Knuth~\cite{knu89}). The Stokes vector of a newly emitted,
unpolarized photon is defined as $\hat{I}_0=(I,Q,U,V)_0^{\rm T}=(1,0,0,0)^{\rm T}$.

Assuming the dust grains to be spherical, the direction of photon reemission
can be determined as for point-like stars shown in Fig.~\ref{emiss}[A] 
(Eq.~\ref{pointlike}). Since the point of reemission inside certain volume elements
of the model space (energy storage cells, ESCs - see Sect.~\ref{esc}) is assumed to be constant,
the location of the reemitting dust grains has to be chosen randomly therein,
assuming a constant density distribution within the cell.
The determination of the point of reemission within an ESC is described for the 
different model geometries shown in Fig.~\ref{gridex} in App.~\ref{escree}.

\subsection{Photon transfer and scattering}\label{ptrsca}

The mean free path length $l$ between the point of emission and the point of the first
photon-dust interaction (and - subsequently - between two points of interaction) is given
by (see App.~\ref{fop})
\begin{equation}\label{tau-ext}
  \tau_{\rm ext,l} = -\ln(1-Z_5),
\end{equation}
\begin{equation}\label{sums}
  \tilde{\tau}_{\rm ext,l}
  = 
  \sum^{i_{\rm end}}_{i=1}
  \left[
  \sum^{n_{\rm D}}_{j=1}
  \rho_j(\vec{r}_{\rm l_i}) \cdot C_{\rm ext_j}
  \right]
  \cdot \Delta l_i,
  \hspace*{1cm} l = \sum^{i_{\rm end}}_{i=1} \Delta l_i.
\end{equation}
Here, $n_{\rm D}$ is the number of different dust grain species and
$\vec{r}_{\rm l_i}$ is the spatial coordinate corresponding to the $i$th
integration point along the path length~$l$.

The step width $\Delta l_i$ has to be small enough to satisfy
the linear approximation of the extinction cross section and density distribution
along the path of the photon (Eq.~\ref{sums}).
The step width $\Delta l_i$ is not fixed but has an initial (maximum) value
which is chosen to be a fraction of the extent of the current energy storage cell 
(ESC, see Sect.~\ref{esc} for an introduction to the ESC concept) the photon
is moving through
-- or, if any of the neighboring cells is smaller, then of that.
This procedure ensures that no ESC is skipped on the photon's path.
Because the subdivision into ESCs is adapted to the model geometry
and distribution of the local optical depth, this coupling drastically increases
the efficiency of the photon transfer since a small step width in high density
regions as well as large step widths in low density (gradient) regions is
provided simultaneously. 
In order to minimize the difference between the exact
value $\tau_{\rm ext,l}$ and the numerically achieved value $\tilde{\tau}_{\rm ext,l}$,
$\Delta l_i$ is decreasing as soon as the test photon reaches the vicinity of the next
interaction point.

This photon transfer concept was found to limit the applicability of previous version of MC3D
to optical depths $\tau \ge 10^{-3}$ (Wolf \& Henning~\cite{wolf00}) since the probability for photons
to interact with the (circumstellar) matter is negligibly small for even lower
optical depths (it decreases exponentially as described by Bouguer-Lambert's law). 
In order to be able to consider lower optical depths,
for instance the simulation of the scattered light arising from an optically thin
envelope or the electron environment around hot stars, the concept of enforced
scattering (Cashwell \& Everett~\cite{ca59}) has been implemented.
The idea is to force each photon to be scattered at least once between
the point of emission and the boundary of the model space, provided there is dust
on its path (optical depth $\tau_{\rm ext_{\rm 0}}$).
A fraction of the photon ($e^{-\tau_{\rm ext_{\rm 0}}}$) leaves the model space
without interaction while the remaining part ($1-e^{-\tau_{\rm ext_{\rm 0}}}$)
will be scattered.
\begin{figure}[t]
 \resizebox{\hsize}{!}{\includegraphics{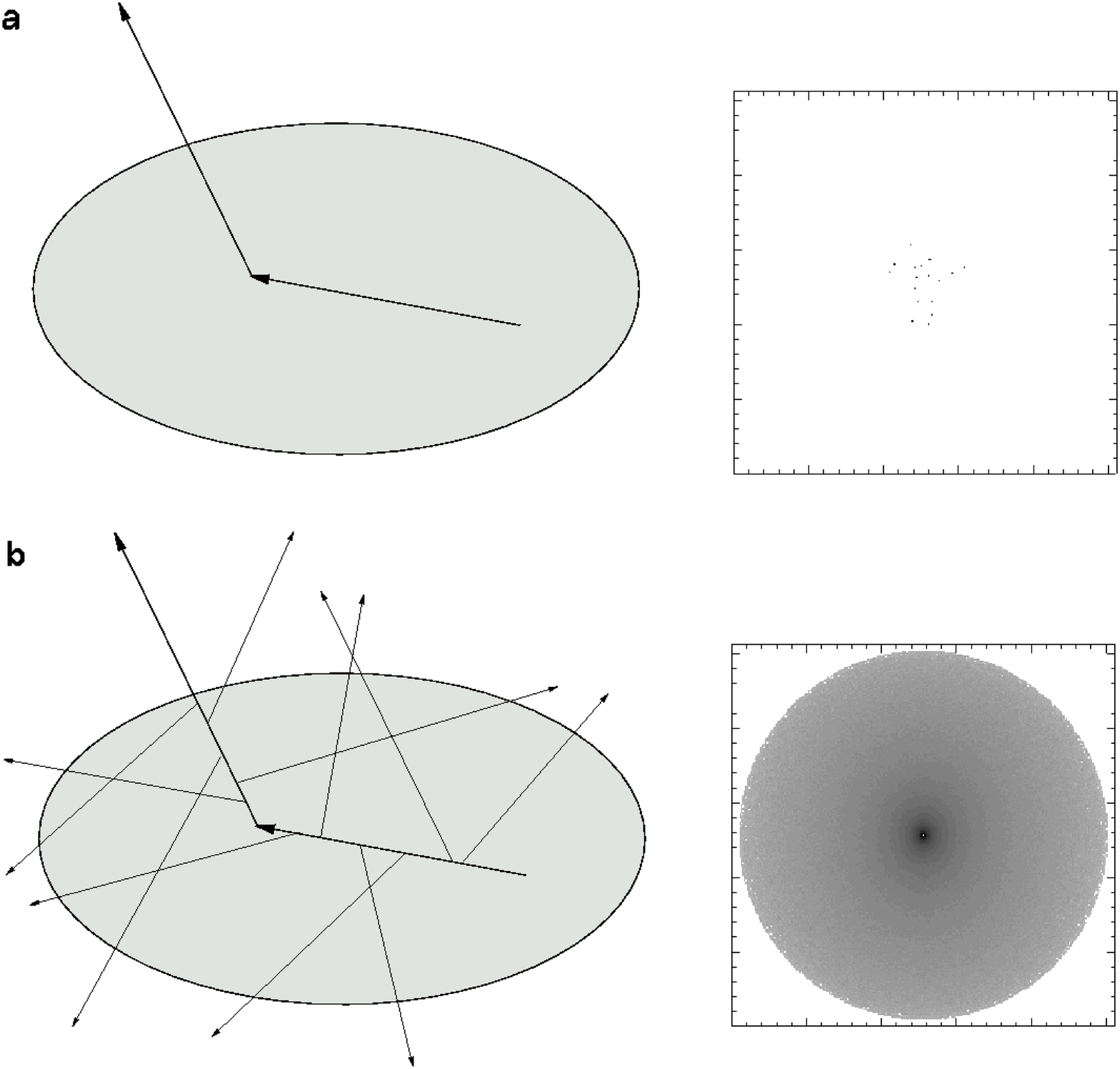}} 
 \caption{Illustration of the high efficiency of the enforced scattering concept in
optically very thin configurations.
(a) Without and (b) with enforced scattering.
Spherical shell being illuminated by an embedded star.
Density profile: $\rho(r) \propto r^{-0.5}$.
Optical depth at the considered wavelength (as seen from the star): $10^{-6}$.
Number of photons: $10^7$. Shown is the (Intensity)$^{1/8}$ without the central star.
The required CPU time amounts in both cases to about 7\,min 
(using a Intel Celeron, 1.2GHz processor).}
 \label{exenfsc}
\end{figure}
Following the same strategy as for the determination of the mean path length
(see App.~\ref{fop}), we derive:
\begin{equation}
  f(\tau_{\rm ext}) {\rm d}\tau_{\rm ext} = 
  \frac{I_{\rm 0} \cdot e^{-\tau_{\rm ext}} \cdot {\rm d}\tau_{\rm ext}}{\int_{0}^{\tau_{\rm ext_{\rm 0}}} I_{\rm 0} \cdot e^{-\tau_{\rm ext}}\, {\rm d}\tau_{\rm ext}} =
  \frac{e^{-\tau_{\rm ext}} \cdot {\rm d}\tau_{\rm ext}}{1-e^{-\tau_{\rm ext_{\rm 0}}}}
\end{equation}
\begin{equation}
  Z 
  = F(\tau_{\rm ext}) 
  = \int_0^{\tau_{\rm ext}}  f(\tau^,_{\rm ext})\, {\rm d}\tau^,_{\rm ext} 
  = \frac{1-e^{-\tau_{\rm ext}}}{1-e^{-\tau_{\rm ext_{\rm 0}}}}
\end{equation}
\begin{equation}
  \tau_{\rm ext} = -\ln{(1-Z[1-e^{-\tau_{\rm ext_{\rm 0}}}])}.
\end{equation}
As Fig.~\ref{exenfsc} illustrates, the enforced scattering concept is of tremendeous
importance for the simulation of scattered light of, e.g., the optically thin
envelope of young stellar objects, debris disks aroung more evolved
stellar systems, or the zodiacal light in extra-solar planetary systems.

Depending on the applied concept for the temperature estimation (see Sect.~\ref{concepts}),
scattering or absorption or both processes occur at a point of interaction.
The modification of the Stokes vector due to the $i$th scattering is described 
by a special M\"uller matrix $\hat{S}_j(\theta,\phi)$
of the dust grain species number $j$
(see, e.g., 
Bickel \& Bailey~\cite{bi85},
Bohren \& Huffman~\cite{bo83}):
\begin{equation} \label{siscsttr}
  \hat{I}_i
  \propto
  \hat{S}_j(\theta,\phi) \hat{I}_{i-1} \ . 
\end{equation}
Here, $\theta$ and $\phi$ are the scattering angles (see App.~\ref{scattmat}).
The estimation of the scattering direction, based on the scattering function
provided by Mie calculations, and the consideration of multiple scattering
events, taking into account (the change of) the polarization state of the weighted
photon, are based on the concepts described by Fischer et al.~\cite{fiwei},
App.~A\,3 and A\,4.2.
The absorption and reemission processes are described in the context of
the temperature estimation in Sect.~\ref{concepts}.

\section{Estimation of the spatial temperature distribution}\label{concepts}

Two different concepts for the solution of the self-consistent RT problem,
i.e., the determination of the spatial dust temperature distribution on the basis
of the amount of absorbed energy of the stellar and the dust reemission 
radiation field, are embedded in MC3D. While the first is based on an iterative
procedure (Sect.~\ref{rtiter}), in the second the temperature distribution is
changed (``updated'') simultaneously with each absorption act (Sect.~\ref{rtimme}).

\subsection{Iterative Radiative Transfer}\label{rtiter}

Assuming a star with an effective temperature $T_{\rm eff}$ and radius $R_*$,
its monochromatic luminosity (assuming blackbody radiation) can be written as
\begin{equation}\label{s_star_e}
L_{\rm \lambda_{\rm, Stern}}^{\rm e} 
= 4 \pi R_*^2 \cdot \pi B_{\rm \lambda}(T_{\rm eff}),
\end{equation}
where $B_{\rm \lambda}(T)$ is the Planck function.
Due to absorption of stellar and reemitted radiation the dust is heated. 
Absorption occurs simultaneously to the scattering whereby the probability for
a dust grain of the species $j$ to be the interaction partner of the photon
is given by
\begin{equation}
\Pi_1(j) =
\frac{\rho_j(\vec{r}) \cdot C_{\rm ext_j}}{\sum_{j'=1}^{n_{\rm D}}\left[ä
\rho_{j'}(\vec{r}) \cdot C_{\rm ext_{j'}}
\right]}
\end{equation}
The difference $\Delta I$  of the Stokes parameter $I$ before and after
the absorption process is determined by the wavelength-dependent
albedo of the dust grains $A_{\rm \lambda_j}$:
\begin{equation}\label{kugabsalb}
  \Delta_i I  = I_{i-1} \cdot ( 1 - A_{\rm \lambda_j} ).
\end{equation}
Consequently, the amount of monochromatic luminosity represented by the particular
photon is decreased by
\begin{equation}\label{absdell}
  \Delta_i L_\lambda = \frac{L_\lambda}{n_{\rm Photon}(\lambda)} \cdot \Delta_i I\ .
\end{equation}

The monochromatic luminosity of a single spherical dust grain with the radius
$a$, the temperature $T_{\rm g}$, and the wavelength-dependent absorption
efficiency $Q^{\rm abs}$ can be written as
\begin{equation}\label{s_grain_e}
L_{\rm \lambda_{\rm, g}}^{\rm e}
= 4 \pi a^2 \cdot Q^{\rm abs}_{\lambda_j}(a)
\cdot \pi B_{\rm \lambda}(T_{\rm g}).
\end{equation}
Assuming a static dust configuration, the energy being reemitted by all grains
of the dust species $j$ within the volume element $V$ during the time interval
$\Delta t$ has to be equal to the energy being absorbed by the grains
(local energy conservation):
\begin{equation} \label{cons_1}
  \Delta t \cdot N_{\rm V_j} \int_0^{\infty} L_{\rm \lambda_j}^{\rm e}\, {\rm d}\lambda
  =
  \Delta t \cdot \int_0^{\infty} L_{\rm \lambda_{\rm V_j}}^{\rm abs}\, {\rm d}\lambda\ ,
\end{equation}
where $N_{\rm V_j}$ is the number of dust grains (species $j$) in the volume $V$.
The absorbed monchromatic luminosity $L_{\rm \lambda_{\rm j, V}}^{\rm abs}$
within the volume $V$ is the accumulated amount of absorbed photon ``fractions'':
\begin{equation}\label{accuphot}
  L_{\rm \lambda_{\rm j, V}}^{\rm abs} = 
  \sum_{k=1}^{n_{\rm Photon}} \sum_{i=0}^{n_{\rm abs}(j,k)} 
  \left( \Delta_i L_{\lambda_j} \right)_k\ ,
\end{equation}
where $n_{\rm abs}(j,k)$ is the number of absorption acts of the $k$th photon
by the dust species $j$ in the volume $V$.
The combination of Eq.~\ref{s_grain_e} and \ref{cons_1} results in an equation
which allows determination of the mean dust grain temperature $\bar{T}_{\rm g_j}$
in the volume $V$:
\begin{equation} \label{tempint}
  \frac{\int_0^{\infty} L_{\rm \lambda_{\rm, V_j}}^{\rm abs}\, {\rm d}\lambda}
  {N_{\rm V} \cdot 4 \pi a^2 \cdot \pi}
  = 
  \int_0^{\infty} Q_{\rm \lambda_j}^{\rm abs}(a) \cdot
  B_{\rm \lambda}(\bar{T}_{\rm g_j})\, {\rm d}\lambda\ .
\end{equation}
Since the temperature $\bar{T}_{\rm g_j}$ cannot be separated from Eq.~\ref{tempint},
in MC3D it is determined from a table of precalculated values of
the right-hand integral in the range of 0\,K to the dust sublimation temperature.
A temperature step width of 0.1\,K was found to be sufficiently small for 
most astrophysical applications.

To consider both, the stellar heating and the subsequent dust reemission radiation,
the following iterative procedure is applied:
\begin{enumerate}
\item Heating by the star.
\item Reemission by the dust grains, based on the energy being absorbed from
the stellar radiation field. The determination of the reemission coordinates
in the ESCs depends on the configuration and is described in App.~\ref{escree}
in detail.
\item Reemission of the dust grains, based on the sum of the energy being
absorbed from the stellar and the dust reemission radiation field.
\end{enumerate}
The third step has to be repeated as long as the difference between the input
energy (given by the stellar radiation field) and the output energy
(given by the sum of the attenuated stellar radiation field and the reemitted
radiation outside the model space) is larger than a user-specific
threshold (e.g., 0.1\% of the input energy).

\subsection{Immediate Reemission}\label{rtimme}

The second concept embedded in MC3D for the estimation of the spatial dust temperature
distribution has been developed by Bjorkman \& Wood~\cite{bj01}.
In contrast to the iterative concept (Sect.~\ref{rtiter}), 
{\sl either} scattering {\sl or} absorption occurs at a point of interaction of a photon
with a dust grain.
The probability for a photon to undergo the one or the other interaction process
is given by
\begin{equation}
\Pi_x(j,\vec{r}) = 
\frac{\rho_j(\vec{r}) \cdot C_{x_j}(\vec{r})}
{\sum_{j'=1}^{n_{\rm D}} \left[\rho_j'(\vec{r}) \cdot C_{ext_j'}(\vec{r})\right]},
\end{equation}
where '$x$' stands for either absorption or scattering.
When a photon packet is absorbed, the new dust temperature is calculated immediately
based on Eq.~\ref{tempint}. To correct the ESC temperature, the wavelength
of the immediately reemitted photon is chosen so that it corrects the temperature
of the spectrum previously emitted by this particular ESC, i.e., the probability
distribution of the reemission wavelength is given by
\begin{equation}
f(\lambda) \propto
Q_{\rm abs}\ 
[B_\lambda(\bar{T}_{\rm g_j}(i)) - B_\lambda(\bar{T}_{\rm g_j}(i-1))],
\end{equation}
where $\bar{T}_{\rm g_j}(i-1)$ represents the temperature of the grain species $j$
before and $\bar{T}_{\rm g_j}(i)$ after the $i$th absorption process in the considered ESC.
The value of the monochromatic luminosity of the reemitted photon results
from the corresponding energy conservation equation:
\begin{equation}
L_{\lambda_1} d\lambda_1 = L_{\lambda_2} d\lambda_2,
\end{equation}
where $d\lambda_1$ and $d\lambda_2$ are the wavelengths of the absorbed/reemitted
photon.
As the simulation runs, the temperature distribution (and the SED given by
the photons leaving the model space -- see Sect.~\ref{obs}) relaxes to its
equilibrium value without iterations.

\begin{figure}[t]
 \resizebox{\hsize}{!}{\includegraphics{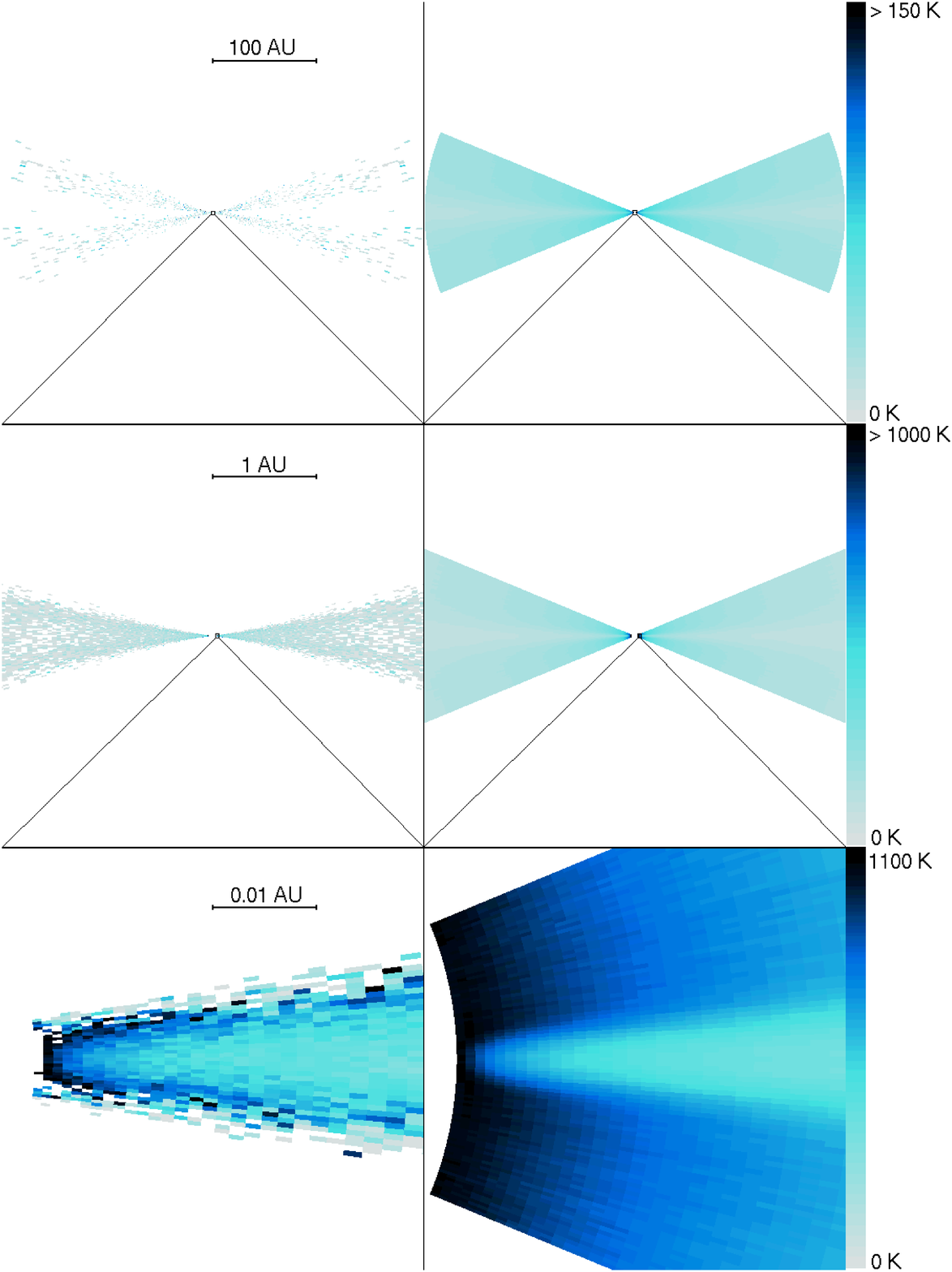}} 
 \caption{Illustration of the high efficiency of the continuous absorption concept.
While the temperature distribution shown in the left column
was simulated without this concept, it was applied to achieve the results 
shown in the right column.
Model: Density distribution of the model for the circumstellar disk around
the T~Tauri star HH\,30 (from Cotera et al.~2001, Wood et al.~2002) but with only
$10^{-2}$ of the mass assumed there. The disk is characterized by an optically
thick midplane and an optically very thin ``atmosphere'' above the disk.
The statistical noise of the temperature values is drastically smaller both in
the regions of high and low optical depth if the continuous absorption
concept is applied (the same number of $10^4$ weighted photons was used
in both simulations).
}
 \label{lucycomp}
\end{figure}

To reduce the number of photons being required for the temperature estimation
in optically thin configurations, the concept of Lucy~\cite{lu99} is applied.
It considers the absorption of the electromagnetic radiation
field not only at the end points of the photon path (points of interaction) but also in between.
The absorption due to all dust grain species has to be taken
into account:
\begin{equation}
\tau_{\rm abs} = \sum_{j=1}^{n_{\rm D}} \int_{\rm point_1}^{\rm point_2} 
C_{\rm abs_j} \cdot \rho_j(\vec{r})\ dr.
\end{equation}
The example of a circumstellar disk with an optically thin envelope shown in 
Fig.~\ref{lucycomp} illustrates the great potential of this concept.

\subsection{ESC concept}\label{esc}

To achieve a spatially resolved dust temperature distribution,
the model space is subdivided into so-called energy storage cells (ESC) in which
the absorbed energy of the radiation field is stored (accumulated) and in which
a constant dust temperature is assumed.
To represent the spatial temperature structure according to the model geometry,
defined by both the spatial distribution of radiation sources and the dust
density, the model space has to be subdivided taking this into account.
Furthermore, the spatial distribution of the optical depth has to be taken into
account in order to resolve the high temperature gradient in optically thick
regions. Finally, the size of the ESCs also has to decrease towards the radiation
source even in case of optically thin configurations to trace the steep
temperature gradient resulting from the geometrical dilution of the radiation field.
Fortunately, the last two conditions are fulfilled simultaneously for most
astrophysical applications since the density gradient increases towards
the radiation source(s).

While the ESC distributions 1D, 2D(a), 2D(b), 3D(b), and 3D(c), shown in 
Fig.~\ref{gridex}, are fully determined by user specific parameters
(such as the number of ESCs and the refinement factors for the model space
towards the central object and/or the disk midplane), model 3D(a)
is prepared to provide an adaptive ESC generation according to
the constraints described above. The model space described by model 3D(a)
is a cube with the side length $l_{\rm c}$. It is subdivided according to
the distribution of the optical depth. Here, the quantity
\begin{equation}
  \rho_{\rm V} = \sum^{n_{\rm D}}_{j=1}
  C_{\rm ext}(j) \cdot \tilde{N}_{\rm V}(j)
\end{equation}
is defined, where $\tilde{N}_{\rm V}(j)$ is the number of dust grains
(species $j$) in the volume element V ($V \le l_{\rm c}^3$). 
At the first step, $\rho_{\rm V}$ is estimated inside the whole model space
($\rho_{\rm V}=\rho_{0}$). In the second step, the model space is subdivided
into 8 cubes (side length $l_{\rm c}/2$). In each ``sub-cube'' the relation
\begin{equation}\label{teilwrel}
  \frac{\rho_{\rm V}(b)}{\rho_{\rm 0}} \le \Xi \le 1
\end{equation}
is checked, whereby $\rho_{\rm V}(b)$ is the value of $\rho_{\rm V}$
in the cube $i_{\rm c}$ and $\Xi$ is a user-specific quantity which controls
the subdivision depth (number of subdivision levels).
In the subsequent steps all sub-cubes for which the relation
(\ref{teilwrel}) is not fulfilled, the subdivision is continued. Finally,
the model space is subdivided into cubes with the side length
$\left(\frac{1}{2}\right)^{i_{\rm c}}l_{\rm c}$ ($i_{\rm c}=1,2,\ldots$).
To prevent this algorithm ending up with too large a number of cubes
(large RAM requirement), resulting from a too small value
given for $\Xi$, the maximum number of cubes is a second user-specific parameter
(constraint) of this algorithm. If the number of cubes exceeds this value,
$\Xi$ is increased, the subdivision process starts again, and so on.

\section{Observables}\label{obs}

Polarization maps, images, and SEDs are the obervables which can be directly
simulated with MC3D.
For this reason the RT transfer is simulated independently  for both the star
and the dust as the sources of emission. The RT is performed as described in
Sect.~\ref{ptrsca} and 
Sect.~\ref{rtiter}, Eq.~\ref{s_star_e}-\ref{s_grain_e} but without iterations.
Although the immediate reemission concept provides the SED and images as a direct result
of the RT too (see Sect.~\ref{rtimme}), it is not discussed here
since the polarization state of the radiation is not considered and a decoupling
of the short-wavelength stellar radiation and the long-wavelength dust reemission
(and the simultaneous decoupling of the radiation at separate wavelengths)
provides more freedom for the selection of the considered wavelength (range)
for which the observables shall be simulated. See also Sect.~\ref{acc}
for a description of the implemented raytracer which can be used to
derive the dust reemission SED and images.

\begin{figure}[t]
 \resizebox{\hsize}{!}{\includegraphics{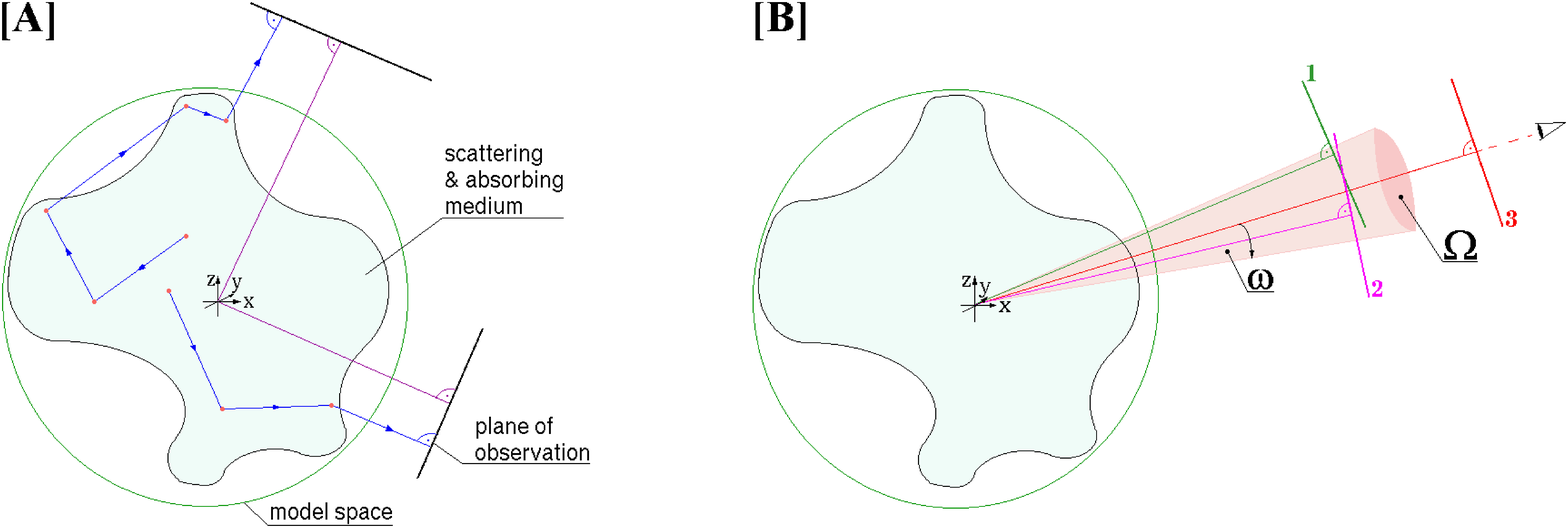}} 
 \caption{{\bf [A]} Random walk and projection of the weighted photons.
{\bf [B]} Definition of the solid angle $\Omega$.
The planes 1 and 2 are co-added with the infinite number of planes
within the solid angle $\Omega$.}
 \label{newrt}
\end{figure}

In order to derive the observables,
a photon will be projected onto an observing plane oriented
perpendicular to the path of the photon as soon as its next scattering
position would be located outside the model space (see Fig.~\ref{newrt}[A]).
Consequently -- in approximation of real observations -- only those photons
will be projected onto the same plane which leave the model space in parallel
directions. The number of observing planes is -- with regard to the possible
last scattering direction -- potentially infinite.

The next step is to combine all those observing maps to a single map
for which the angular distance of their normal vector to a given direction
is smaller than the angle $\omega$ (see Fig.~\ref{newrt}[B]).
The sum of the monochromatic luminosity of all photons collected on this map
therefore represents the monochromatic luminosity 
$L_{\lambda,\Omega}'$ ($L_{\lambda,\Omega}' \le L_{\lambda}'$)
of the object seen under the solid angle $\Omega_{\rm [sr]}=2\pi(1-\cos\omega_{\rm [rad]})$.

Let $\sigma_{\rm A}$ be the radiating area of an object seen in projection
on the observing plane.
Then, the corresponding monochromatic intensity $I_{\lambda}$ of the object is
\begin{equation}
  I_{\lambda} = \frac{L_{\lambda, \Omega}'}{\Omega \cdot \sigma_{\rm A}}\ .
\end{equation}
The flux density $S_{\lambda}$ at the point of observation is determined by the 
distance $d$ between the object and the observer as follows:
\begin{equation}
  S_{\lambda} = \frac{\sigma_{\rm A}}{d^2} \cdot I_{\lambda}\ .
\end{equation}
In order to derive spatially resolved images (such as shown in Fig.~\ref{lucycomp}),
the observing planes are subdivided in pixels. Regarding the last scattering
position of a photon, its (projected) Stokes parameters, wavelength and monochromatic
luminosity are stored in the corresponding pixel. The required transformation
of the Stokes vector was outlined by Fischer~\cite{fiphd}.

\section{Diverse further embedded concepts}\label{acc}

\begin{enumerate}
\item Selective Emission (for self-consistent RT)
  \begin{enumerate}
  \item {\em Wavelength range selection:}\\
    The CPU time can be drastically reduced by considering only those wavelength ranges
    in which the particular radiative source (the star or the dust in a certain ESC) is
    emitting a non-negligible amount of energy (see Wolf \& Henning~2000) for details.
  \item {\em Geometrical selection:}\\
    For certain astrophysical models, the efficiency can be improved by excluding
    photons which would be emitted into dust free regions.\\
    {\sl \underline{Example:} 
      Circumstellar disk with a very small opening angle $\xi$
      (see Fig.~\ref{gridex}, models 2D(a) and 2D(b)) around a pointlike star:
      Considering only those photons being emitted into the disk, the number of photons
      required to achieve a certain accuracy is reduced by a factor of $1/\sin{\xi}$.
      In contrast to Eq.~\ref{pointlike}, the angles of emission are now given by
      \begin{equation}
	\cos\theta_{\rm E} = (-1 + 2\, Z_6) \cdot \sin(\xi), \hspace*{1cm} 
	\phi_{\rm E}       = 2\,\pi\,Z_7.
      \end{equation}
    }
  \end{enumerate}
\item Taking into account the symmetry of the considered model, the observing concept 
described in Sect.~\ref{obs} can be modified in order to combine observing planes to a larger
(effective) solid angle $\Omega$. While in the case of 1D symmetry the angle $\omega$ can be
chosen to be 180$^{\rm o}$, in models with the z axis as an axis of rotation symmetry,
all planes within the angular region $(\theta_0-\omega) \le \theta \le (\theta_0+\omega)$
are combined, whereby $\theta_0$ is the angle between the $z$ axis and the normal vector
of the observing plane. The corresponding (effective) solid angle $\Omega$ which 
is required to normalize the flux/intensity then amounts to 
$\Omega = 2\pi [\cos(\theta_0-\omega) - \cos(\theta_0+\omega)]$.

\item MC3D is prepared to take into account initial dust temperatures $T_{\rm g} > 0$,
such as in active accretion disks around young stellar objects. Therefore,
the following modifications to the self-consistent RT process have to be applied:
\begin{itemize}
\item {\em Iterative reemission:}
The initial dust radiation field resulting from the initial dust temperature
and dust density distribution (and dust grain optical properties) is added to
the absorbed stellar radiation field in Eq.~\ref{cons_1}.
\item {\em Immediate concept:}
The dust temperatures are set before the stellar RT is started.
Thus, the absorbed stellar photons provide an additional increase of the dust grain
temperature.
\end{itemize}

\item MC3D is equipped with a raytracer being optimized for the 1D and 2D models shown
in Fig.~\ref{gridex}. It allows derivation of images and the SED resulting from the dust reemission
radiation. The basing assumption that scattering processes are negligible in the
corresponding mid-infrared-millimeter wavelength range is very well fulfilled
for most of the astrophysical applications targeted by MC3D.

\begin{figure}[t]
 \resizebox{\hsize}{!}{\includegraphics{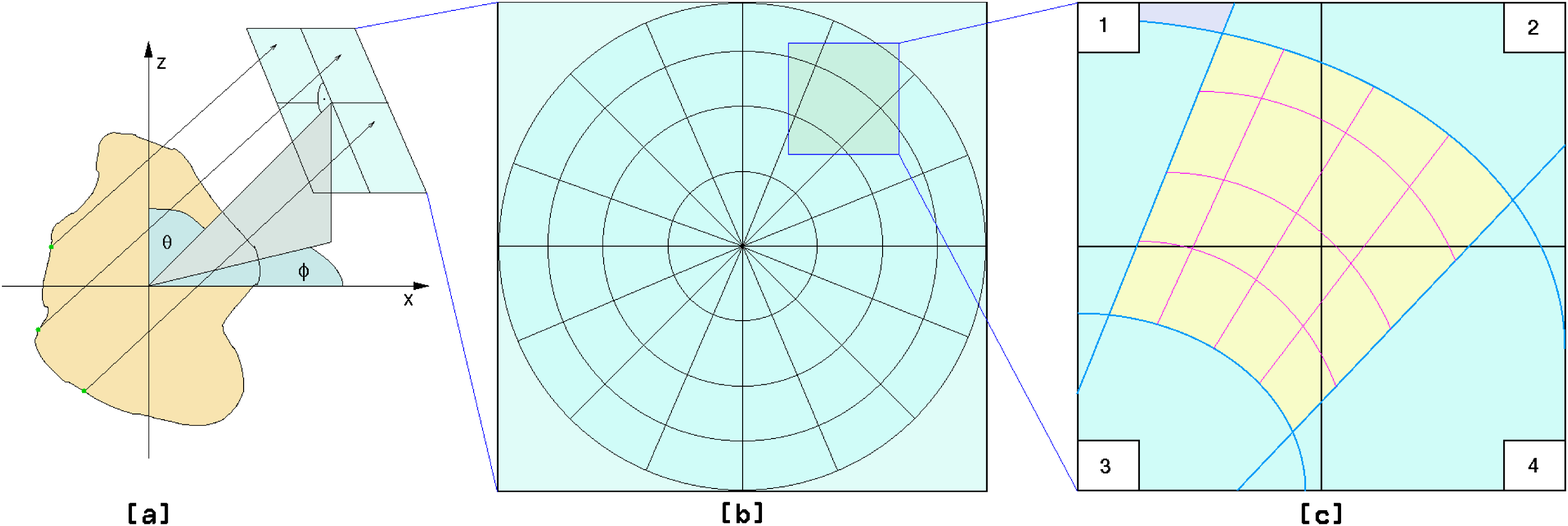}} 
 \caption{Raytracer. 
{\bf [a]} Projection of the monochromatic luminosity on the observing plane.
{\bf [b]} Configuration of the ray location on the observing plane.
{\bf [c]} Projection of each mapped region (represented by single ray) onto the
matrix-like observing plane. The squares 1-4 symbolize single pixels.}
 \label{rtrace}
\end{figure}
Along rays, starting opposite to the observer's location (with respect to the model space),
the contribution to the monochromatic luminosity is integrated taking into account
extinction (see Fig.~\ref{rtrace}[A]).
To take into account the emission/extinction of the smallest structures but
also to guarantee a fast integration process, the step width of the numerical integration 
along a ray is adapted to the ESC grid, using a fraction of the ESC of the current 
integration point. The starting points of the rays are located in centrosymmetric
rings (see Fig.~\ref{rtrace}[B]), whereby the inner/outer radii of these rings correspond
to the radial subdivision of the model space into ESCs. The area whithin the inner
radius of the models (where no subdivision of the model into ESCs is provided)
is covered by rings with the inner/outer radius difference equal to that of the innermost ESC(s).
Furthermore, the rings are subdivided in the azimuthal direction whereby the step width
is chosen to achieve a similar resolution as in the radial direction.
To increase the spatial resolution of the maps, each ring can be subdivided 
in a certain number of rings, and consequently smaller azimuthal subdivisions.
To project the monochromatic luminosity determined in each of the resulting segments (each represented
by a single ray) as shown in Fig.~\ref{rtrace}[B] and [C] on the matrix-like observing
map, they are subdivided in radial and azimuthal direction again.
\end{enumerate}

\section{Acknowledgements}\label{ackno}
I gratefully thank Olaf Fischer for his introduction into Monte Carlo radiative transfer
-- the numerical implementation of the
basic scattering mechanism by spherical grains is mainly based on
algorithms developed by him (Fischer~\cite{fiphd}).
I wish to thank both anonymous referees for their careful reading
of the manuscript and their valuable comments and suggestions.
This research was supported through the HST Grant GO\,9160
through the NASA grant NAG5-11645,
and through the DFG grant Ste 605/10 within the program
``Physics of star formation''.

\newpage
\begin{appendix}

\section{Weighted mean optical parameters of dust grain mixtures}\label{mix}

In case of non-self-consistent radiative transfer (RT) in a dust grain mixture,
the numerical effort (ultimately, the CPU time and RAM requirement) may be substantially
decreased by assuming weighted mean values of those parameters which describe
the interaction of the electromagnetic field with the dust grains.
The weight $w_j(a)$ which represents the contribution of a certain component
of the dust grain mixture results from the abundance of the $i$th material
of given dust grain number density (assuming $n_{\rm D}$ chemically
different dust species) and the size distribution of the respective material:
\begin{equation}\label{weight}
 \sum^{n_{\rm D}}_{i=1}\int^{a_{\rm max}}_{a_{\rm min}} w_j(a)\, da = 1\ .
\end{equation}
The Stokes parameters as well as the extinction, absorption, and scattering
cross section ($C_{\rm ext}$, $C_{\rm abs}$, $C_{\rm sca}$) are additive. 
Therefore, the representative values in case
of a dust grain mixture can be derived on the basis of their weighted contributions
(see, e.g.,  Martin~1978, \u{S}olc~1980):
\begin{equation}\label{cextmean}
  \langle C_{\rm ext} \rangle = \sum^{n_{\rm D}}_{i=j} \int^{a_{\rm max}}_{a_{\rm min}} 
  w_j(a) \cdot C_{{\rm ext}_j}(a)\, da,
\end{equation}
\begin{equation}
  \langle C_{\rm abs} \rangle = \sum^{n_{\rm D}}_{i=j} \int^{a_{\rm max}}_{a_{\rm min}} 
  w_j(a) \cdot C_{{\rm abs}_j}(a)\, da,
\end{equation}
\begin{equation}
  \langle C_{\rm sca} \rangle = \sum^{n_{\rm D}}_{i=j} \int^{a_{\rm max}}_{a_{\rm min}} 
  w_j(a) \cdot C_{{\rm sca}_j}(a)\, da,
\end{equation}
and
\begin{equation}
  \langle \hat{S} \rangle = \sum^{n_{\rm D}}_{i=j} \int^{a_{\rm max}}_{a_{\rm min}} 
  w_j(a) \cdot \hat{S}_j(a)\, da,
\end{equation}
where $\hat{S}$ is the M\"uller matrix which is used to describe the modification
of the photon's Stokes vector due to the interaction of a photon with the 
scattering/absorbing medium (dust grains; see Bickel \& Bailey~1985, Bohren \& Huffman~1983).
For the albedo $A$ follows:
\begin{equation}\label{albmean}
  \langle A \rangle = \frac{\sum^{n_{\rm D}}_{j=1}\int^{a_{\rm max}}_{a_{\rm min}}
    w_j(a) \cdot C_{{\rm ext}_j}(a) \cdot A_j(a)\, da}
  {\sum^{n_{\rm D}}_{i=1}\int^{a_{\rm max}}_{a_{\rm min}}
    w_j(a) \cdot C_{{\rm ext}_j}(a)\, da}
  =\frac{\langle C_{\rm sca} \rangle}{\langle C_{\rm ext} \rangle}.
\end{equation}

\section{Mean free path length}\label{fop}

\begin{enumerate}
\item Decrease of the intensity of a plane wave due to extinction,
described by Bouguer-Lambert's law
  \begin{equation}\label{bouglamb}
    I_{\rm 1} = I_{\rm 0} \cdot e^{-\tau}
  \end{equation}
  ($I_{\rm 0}\,\ldots$\,original intensity, 
  $I_{\rm 1}\,\ldots$\, intensity after passing a medium with the optical depth $\tau$).

\item Let $\tau_{\rm ext}$ be the optical depth due to dust extinction:
  \begin{equation}
    \tau = \tau_{\rm ext} = 
  \left[
  \sum^{n_{\rm D}}_{j=1} 
  C_{\rm ext_j} \cdot \rho_j 
  \right]
  \cdot s
  \end{equation}
  ($C_{\rm ext_j}$ is the wavelength-dependent extinction cross section of a single 
  dust grain of the species number $j$,
  $s$ is the geometric free path length,
  $\rho_j$ is the number density of dust particles of that species)\\
  assumption: 
  $C_{\rm ext_j} \cdot \rho_j$ is constant along the geometrical distance $s$ 

\item Probability $f(\tau_{\rm ext}) {\rm d}\tau_{\rm ext}$ for a photon to pass
  a distance corresponding to the optical depth $\tau_{\rm ext}$:
  \begin{equation}\label{ofwvot}
  \begin{array}{l}
    f(\tau_{\rm ext}) {\rm d}\tau_{\rm ext} = 
    \frac{I_{\rm 0} \cdot e^{-\tau_{\rm ext}} \cdot {\rm d}\tau_{\rm ext}}
    {\int_{0}^{\infty} I_{\rm 0} \cdot e^{-\tau_{\rm ext}}\,{\rm d}\tau_{\rm ext}}\\
    = \frac{I_{\rm 0} \cdot e^{-\tau_{\rm ext}} \cdot {\rm d}\tau_{\rm ext}}{I_{\rm 0}}\\
    = e^{-\tau_{\rm ext}} \cdot {\rm d}\tau_{\rm ext}
  \end{array}
  \end{equation}

\item Resulting distribution of the optical depth $F(\tau_{\rm ext})$:
  \begin{equation}
    F(\tau_{\rm ext}) = \int_{0}^{\tau_{\rm ext}} e^{-\tau_{\rm ext}^,}\, {\rm d}\tau_{\rm ext}^, 
  \end{equation}

\item Determination of the free optical path langth by substitution of
$F(\tau_{\rm ext})$ by a random number uniformly distributed in the intervall $[0,1]$:
  \begin{equation}
    \tau_{\rm ext} = -\ln{(1-Z)}\ .
  \end{equation}
\end{enumerate}

\section{Scattering matrix}\label{scattmat}

The scattering Matrix (special M\"uller matrix) for the description of the modification
of the Stokes vector due to the interaction of a weighted photon with a spherical,
homogeneous dust particle (Mie scattering) has the form
\begin{equation}\label{kugscmat}
  \hat{S}(\theta) =
  \left(
    \begin{array}{cccc}
      S_{11}(\theta) & S_{12}(\theta) &  0              & 0\\
      S_{12}(\theta) & S_{11}(\theta) &  0              & 0\\
      0              & 0              &  S_{33}(\theta) & S_{34}(\theta)\\
      0              & 0              & -S_{34}(\theta) & S_{33}(\theta)
    \end{array}
  \right),
\end{equation}
where
\xbeq
\begin{array}{l}
S_{11}(\theta)  =  \frac{1}{2}( \left| S_1(\theta) \right|^2   
                            +  \left| S_2(\theta) \right|^2   
                            +  \left| S_3(\theta) \right|^2   
                            +  \left| S_4(\theta) \right|^2 )\\
S_{12}(\theta)  =  \frac{1}{2}( \left| S_2(\theta) \right|^2 
                            -  \left| S_1(\theta) \right|^2   
                            +  \left| S_4(\theta) \right|^2  
                            -  \left| S_3(\theta) \right|^2 )\\
S_{33}(\theta)  =    {\rm Re}\{S_1(\theta) S^*_2(\theta) 
                   + S_3(\theta) S^*_4(\theta)\} \\
S_{34}(\theta)  =    {\rm Re}\{S_2(\theta) S^*_1(\theta) 
                   + S_4(\theta) S^*_3(\theta)\}
\end{array}
\xeeq
\xbeq
\begin{array}{l}
S_{\rm 1} = \sum^{\infty}_{n=1} \frac{2n+1}{n(n+1)}(a_n\pi_n  + b_n\tau_n),
\\
S_{\rm 2} = \sum^{\infty}_{n=1} \frac{2n+1}{n(n+1)}(a_n\tau_n + b_n\pi_n),
\\
S_{\rm 3} = S_{\rm 4} = 0
\end{array}
\xeeq
\xbeq
\begin{array}{l}
a_n = \frac{\psi_n'(mx)\psi_n(x)  - m\psi_n(mx)\psi_n'(x)}
           {\psi_n'(mx)\zeta_n(x) - m\psi_n(mx)\zeta_n'(x)},\ 
\\
\\
b_n = \frac{m\psi_n'(mx)\psi_n(x) - \psi_n(mx)\psi_n'(x)}
           {m\psi_n'(mx)\zeta_n(x) - \psi_n(mx)\zeta_n'(x)}
\end{array}
\xeeq
\xbeq
\begin{array}{l}
\pi_n(\cos\theta)  = \frac{1}{\sin\theta}P^1_n(\cos\theta),\ \hspace*{0.5cm}
\\
\\
\tau_n(\cos\theta) = \frac{d}{d\theta}P^1_n(\cos\theta)\ .
\end{array}
\xeeq
$P^1_n(\cos\theta)$ are associated Legendre functions,
$\psi_n(x)$ und $\zeta_n(x)$ are Riccati-Bessel functions,
$m$ is the relative refraction index, and
\begin{equation}\label{grpar}
  x = \frac{2\pi}{\lambda} \cdot \|m_0\| \cdot a_{\rm K}
\end{equation}
is the size parameter. The quantity $m_0$ is the refractive index of the surrounding
medium\footnote{Since MC3D was developed to model astrophysical objects for which
the surrounding medium can be assumed to be a perfect vacuum, $\|m_0\|$=1.}
$\lambda$ is the wavelength of the infalling radiation and $ a_{\rm K}$
is the dust particle radius (see Bohren \& Huffman~\cite{bo83}).

The numerical formulation of the scattering mechanism for spherical,
homogenous particle is based on the algorithm presented by Fischer~\cite{fiphd}.

In case of Thomson scattering, the scattering matrix elements are 
wavelength-independent and can be written as
\begin{equation}
  \begin{array}{l}
    S_{\rm 11}(\theta) = S_{\rm 22}(\theta) = (\cos^2(\theta)+1)/2\\
    S_{\rm 12}(\theta) = S_{\rm 21}(\theta) = (\cos^2(\theta)-1)/2\\
    S_{\rm 33}(\theta) = S_{\rm 44}(\theta) = \cos(\theta)\\
    S_{\rm 13}=S_{\rm 31}=S_{\rm 23}=S_{\rm 32}=0\\
    S_{\rm 14}=S_{\rm 24}=S_{\rm 34}=S_{\rm 43}=0\\
  \end{array}
\end{equation}
(see Bohren \& Huffman~\cite{bo83}).

The scattering cross section of an electron in this scattering regime is given by
\begin{equation}
  \sigma_{\rm T} = \frac{8\pi}{3} \cdot 
  \left( \frac{e^2}{4 \pi \varepsilon_{\rm 0} m_{\rm e} c^2} \right)^2
\end{equation}
($e$ is the elementary charge, 
$\varepsilon_{\rm 0}$ is the dielectric constant of free space;
see, e.g., Musiol et al.~\cite{mu88}).

\section{Reemission from ESCs}\label{escree}

In the following the equations for the (random) determination of
the reemission point within an ESC for the different models shown in Fig.~\ref{gridex}
are given. According to the symmetry of the particular model,
spherical $(r,\theta,\phi)$,
cylindrical $(r,\phi,z)$, or
cartesian coordinates ($x,y,z$) are used.

{\sl \small
Definition:
Let $\varepsilon$ be $\epsilon \ \{x,y,z,r,\phi,\theta\}$. 
Then, $\varepsilon_{i}$ and $\varepsilon_{i+1}$ are the inner and outer boundary
of the ESC along the coordinate $\varepsilon$, whereby $\varepsilon_{i+1} > \varepsilon_{i}$.
}

\setlongtables
\begin{longtable}{p{4cm}p{10cm}}
\hline

{\bf Model: 1D}
\newline
\begin{center}
Definition ranges:\newline
$R_{\rm in} \le r      \le R_{\rm ou} $\newline
$0^{\rm o}  \le \theta \le 180^{\rm o}$\newline
$0^{\rm o}  \le \phi   <   360^{\rm o}$\newline
\end{center}
&
\vspace*{-2mm}

\begin{equation}
r = \left[ Z_1 (r_{i+1}^3 - r_i^3) + r_i^3 \right]^{\frac{1}{3}}
\end{equation}

\begin{equation}
\cos{\theta} = 1 - 2 Z_2,
\end{equation}

\begin{equation}
\phi = 2 \pi Z_3
\end{equation}
\\
\hline
{\bf Model: 2D(a) }
\newline
\begin{center}
Definition ranges:\newline
$R_{\rm in} \le r      \le R_{\rm ou}        $\newline
$\xi        \le \theta \le 180^{\rm o} - \xi $\newline
$0^{\rm o}  \le \phi   <   360^{\rm o}       $\newline
\end{center}
&
\vspace*{-2mm}

\begin{equation}
r = \left[ Z_1  (r_{i+1}^3 - r_i^3) + r_i^3 \right]^{\frac{1}{3}}
\end{equation}

\begin{equation}
\cos{\theta} = (1 - 2 Z_2) \sin{\xi}
\end{equation}

\begin{equation}
\phi = 2 \pi Z_3
\end{equation}
\\
\hline
{\bf Model: 2D(b)}
\newline
\begin{center}
Definition ranges:\newline
$R_{\rm in} \le r      \le R_{\rm ou}        $\newline
$\xi        \le \theta \le 180^{\rm o} - \xi $\newline
$0^{\rm o}  \le \phi   <   360^{\rm o}       $\newline
\end{center}
&
\vspace*{-2mm}

\begin{equation}
r = \left[ Z_1 (r_{i+1}^3 - r_i^3) + r_i^3 \right]^{\frac{1}{3}}
\end{equation}

\begin{equation}
\cos{\theta} = \cos{\theta_{i+1}} + Z_2 (\cos{\theta_{i}-\cos{\theta_{i+1}}})
\end{equation}

\begin{equation}
\phi = 2 \pi Z_3
\end{equation}
\\
\hline
{\bf Model: 3D(a)}
\newline
\begin{center}
Definition ranges:\newline
$0 \le x      \le l_{\rm c}        $\newline
$0 \le y      \le l_{\rm c}        $\newline
$0 \le z      \le l_{\rm c}        $\newline
\end{center}
&
\vspace*{-2mm}

\begin{equation}
x = x_i + Z_1 (x_{i+1}-x_{i})
\end{equation}

\begin{equation}
y = y_i + Z_2 (y_{i+1}-y_{i})
\end{equation}

\begin{equation}
z = z_i + Z_3 (z_{i+1}-z_{i})
\end{equation}
\\
\hline
{\bf Model: 3D(b)}
\newline
\begin{center}
Definition ranges:\newline
$R_{\rm in} \le r      \le R_{\rm ou}        $\newline
$0^{\rm o}  \le \theta \le 180^{\rm o}       $\newline
$0^{\rm o}  \le \phi   <   360^{\rm o}       $\newline
\end{center}
&
\vspace*{-2mm}

\begin{equation}
r = \left[ Z_1 (r_{i+1}^2 - r_i^2) + r_i^2 \right]^{\frac{1}{2}}
\end{equation}

\begin{equation}
\phi = \phi_i + Z_2 (\phi_{i+1}-\phi_i)
\end{equation}

\begin{equation}
z = z_i + Z_3 (z_{i+1} - z_i)
\end{equation}
\\
\hline
{\bf Model 3D(c)}
\newline
\begin{center}
Definition ranges:\newline
$R_{\rm in} \le r      \le R_{\rm ou}        $\newline
$0^{\rm o}  \le \theta \le 180^{\rm o}       $\newline
$0^{\rm o}  \le \phi   <   360^{\rm o}       $\newline
\end{center}
&
\vspace*{-2mm}

\begin{equation}
r = \left[ Z_1 (r_{i+1}^3 - r_i^3) + r_i^3 \right]^{\frac{1}{3}}
\end{equation}

\begin{equation}
\cos{\theta} = \cos{\theta_{i}} + Z_2 (\cos{\theta_{i}}-\cos{\theta_{i+1}})
\end{equation}

\begin{equation}
\phi = \phi_i + Z_3 (\phi_{i+1}-\phi_{i})
\end{equation}
\\
\hline
\end{longtable}



\end{appendix}
\end{document}